\documentclass[aps,prl,reprint,superscriptaddress,nofootinbib,floatfix]{revtex4-2}
\usepackage{amsfonts,amssymb,amsmath}
\usepackage[dvipsnames, table]{xcolor}
\usepackage{epsfig}
\usepackage{bm}
\usepackage{latexsym}
\usepackage{natbib}
\usepackage{url}
\usepackage{dcolumn}
\usepackage{color}
\usepackage{graphicx,epsfig}
\usepackage{comment}
\usepackage{hyperref}
\usepackage[normalem]{ulem}
\hypersetup{colorlinks   = true,
            urlcolor     = blue,
            citecolor    = blue,
            linkcolor    = blue,
            menucolor    = blue,
            anchorcolor  = blue,
            filecolor    = blue
}

\begin{document}
\title{Testing the dark origin of neutrino masses with oscillation experiments}

\newcommand{\UniSA}{\affiliation{Dipartimento di Fisica ``E.R.\ Caianiello'', Universit\`a degli Studi di Salerno,\\ Via Giovanni Paolo II, 132 - 84084 Fisciano (SA), Italy}}
\newcommand{\INFN}{\affiliation{Istituto Nazionale di Fisica Nucleare - Gruppo Collegato di Salerno - Sezione di Napoli,\\ Via Giovanni Paolo II, 132 - 84084 Fisciano (SA), Italy}}
\newcommand{\TDLI}{\affiliation{Tsung-Dao Lee Institute \& School of Physics and Astronomy, Shanghai Jiao Tong University, Shanghai 201210, China}}

\author{Andrew Cheek}
\email{acheek@sjtu.edu.cn}
\TDLI
\author{Luca Visinelli}
\email{lvisinelli@unisa.it}
\UniSA \INFN
\author{Hong-Yi Zhang}
\email{hongyi18@sjtu.edu.cn \newline }
\TDLI

\date{\today}
\begin{abstract}
The origin of neutrino masses remains unknown to date. One popular idea involves interactions between neutrinos and ultralight dark matter, described as fields or particles with masses $m_\phi \ll 10\,\mathrm{eV}$. Due to the large phase-space number density, this type of dark matter exists in coherent states and can be effectively described by an oscillating classical field. As a result, neutrino mass-squared differences undergo field-induced interference in spacetime, potentially generating detectable effects in oscillation experiments. We demonstrate that if $m_\phi\gg 10^{-14}\,\mathrm{eV}$, the mechanism becomes sensitive to dark matter density fluctuations, which suppresses the oscillatory behavior of flavor-changing probabilities as a function of neutrino propagation distance in a model-independent way, thereby ruling out this regime. Furthermore, by analyzing data from the Kamioka Liquid Scintillator Antineutrino Detector (KamLAND), a benchmark long-baseline reactor experiment, we show that the hypothesis of a dark origin for the neutrino masses is disfavored for $m_\phi \ll 10^{-14}\,\mathrm{eV}$, compared to the case of constant mass values in vacuum. This result holds at more than the 4$\sigma$ level across different datasets and parameter choices. The mass range $10^{-17}\,\mathrm{eV} \lesssim m_\phi \lesssim 10^{-14}\,\mathrm{eV}$ can be further tested in current and future oscillation experiments by searching for time variations (rather than periodicity) in oscillation parameters.
\end{abstract}

\maketitle

{\bf \emph{Introduction.}}---Despite the success of the Standard Model of particle physics in explaining a wide range of phenomena, two key observations demand physics beyond it. The first is neutrino oscillations, which proves that neutrinos have masses and undergo flavor transitions as they propagate~\cite{Pontecorvo:1967fh}. The proposed solutions typically involve new mass scales or additional particles, such as the seesaw mechanism~\cite{Yanagida:1979as, Gell-Mann:1979vob, Mohapatra:1979ia, Schechter:1980gr}. The second is the evidence for dark matter, which dominates the matter content of the present-day Universe~\cite{Trimble:1987ee, Planck:2018nkj}.

It is tempting to speculate that these two puzzles emerge from an interconnected physical origin. For example, neutrino masses can be generated by interactions with dark matter through a long-range scalar force~\cite{Davoudiasl:2018hjw}. In addition, neutrino forward scattering on background dark matter particles could yield a neutrino potential with a dependence on the neutrino energy $E_\nu$ as $\sim (2 E_\nu)^{-1}$, potentially explaining neutrino oscillations~\cite{Ge:2018uhz, Choi:2019zxy, Sen:2023uga, Capozzi:2018bps}. Small neutrino masses can also be generated through radiative symmetry breaking \cite{Lee:2024rdc}. In the cases where the dark matter field couples to sterile neutrinos that share the exotic forces with light active neutrinos by mixing~\cite{Huang:2022wmz, Plestid:2024kyy}, it may even induce oscillations between the Dirac and Majorana nature of neutrinos~\cite{ChoeJo:2023ffp,Kim:2025xum}. It is recently suggested to test the dark origin of neutrino masses by searching for supernova neutrinos with a galactic origin~\cite{Ge:2024ftz}. Apart from masses, the dark origin of mixing angles is also considered in~\cite{Liao:2018byh}. 

If generated from interactions with scalar dark matter, neutrino masses $m_i$ for the mass eigenstate $i$ would depend on the local galactic dark matter density $\rho$ through~\cite{Ge:2018uhz, Choi:2019zxy, Sen:2023uga, Capozzi:2018bps, Lee:2024rdc, Huang:2022wmz, Plestid:2024kyy, ChoeJo:2023ffp}
\begin{align}
\label{neutrino_mass}
m_i \sim 0.2 \,\mathrm{eV} \left( \frac{g_i}{10^{-7}} \right) \left( \frac{10^{-9} \,\mathrm{eV}}{m_\phi} \right) \left( \frac{\rho}{0.4 \,\mathrm{GeV}/\mathrm{cm}^3} \right)^{1/2} ~,
\end{align}
where $g_i$ is the effective neutrino-scalar coupling, and $m_\phi$ is the mass of dark matter. Given the current constraint from supernova cooling $g_i \lesssim 10^{-7}$~\cite{Farzan:2018gtr}, the required mass of dark matter should satisfy $m_\phi\lesssim 10^{-9}\,\mathrm{eV}$ to account for a dominant contribution to neutrino masses~\cite{Sen:2023uga}.\footnote{Neutrinos with a bare mass $m_i$ can induce loop corrections to $m_\phi$. For instance, the one-loop radiative correction due to a Yukawa coupling is approximately $\Delta m_{\phi}^2\sim g_i^2 m_i^2 / (8\pi^2)$. To maintain a small $m_\phi$ without fine-tuning, the coupling constant must satisfy $g_i\lesssim 10^{-7}\left(m_\phi/10^{-9}\,{\rm eV}\right)\left(0.1\,{\rm eV}/m_i\right)$. Since this constraint is model-dependent, we do not elaborate further.} On the other hand, the lower bound of the dark matter mass, inferred from the dynamical heating of stars in dwarf galaxies and the matter power spectrum, is given by $m_\phi \gtrsim 3\times 10^{-19} \,\mathrm{eV}$ \,\cite{Dalal:2022rmp, Amin:2022nlh}.

In this mass range, dark matter has a large average state occupation number~\cite{Hui:2021tkt, Ferreira:2020fam}
\begin{align}
n \lambda_\mathrm{dB}^3 \simeq \left( \frac{\rho}{0.4\, \mathrm{GeV}/\mathrm{cm}^3} \right) \left( \frac{40 \,\mathrm{eV}}{m_\phi} \right)^4 \left( \frac{200\mathrm{km}/\mathrm{s}}{v} \right)^3 ~,
\end{align}
where $n=\rho/m_\phi$ is the dark matter number density, $\lambda_\mathrm{dB}$ is the de Broglie wavelength, and $v$ is the typical velocity dispersion in the Milky Way galaxy. States with a large occupation number $n\lambda_\mathrm{dB}^3\gg 1$ are well described as coherent states, as a result the dark matter field can be viewed as an oscillating classical scalar field $\phi \propto \cos(m_\phi t - \boldsymbol v \cdot \boldsymbol x)$~\cite{Dvali:2017ruz, Allali:2021puy, Ferreira:2020fam, Guth:2014hsa, Hui:2021tkt}. The resulting neutrino mass-squared difference undergoes field interference in spacetime, leading to distorted neutrino oscillation probabilities~\cite{Krnjaic:2017zlz, Brdar:2017kbt, Dev:2020kgz}.

In this {\it Letter}, we investigate the dark origin of neutrino masses due to ultralight scalar dark matter through neutrino oscillation experiments. The interference of oscillation probabilities in time and space allows us to impose strong constraints on this scenario for all mass ranges. The range $10^{-17} \,\mathrm{eV} \lesssim m_\phi \lesssim 10^{-14} \,\mathrm{eV}$ can be further constrained in current and future oscillation experiments in time scales from 10 days to a few decades.

{\bf \emph{Neutrino oscillation probabilities.}}---In the presence of an oscillating classical scalar field as dark matter and for nonuniversal couplings $g_i\neq g_j$ for $i\neq j$, the neutrino mass-squared difference can be parameterized as~\cite{Choi:2019zxy, Sen:2023uga, Capozzi:2018bps, Lee:2024rdc, Huang:2022wmz, Plestid:2024kyy, ChoeJo:2023ffp}
\begin{align}
    \label{mass_squared_difference}
    \Delta m_{ij}^2 = \Delta m_{ijD}^2(\boldsymbol x) \cos^2(m_\phi t) ~,
\end{align}
where $\Delta m_{ijD}^2 \propto 2\rho(\boldsymbol x) / m_\phi^2$ is the oscillation amplitude of the mass-squared difference, where the subscript ``$D$'' emphasizes its dark origin. We neglect the space-dependent phase in the cosine function, since it remains a small factor during neutrino propagation~\cite{Berlin:2016woy}. To interpret this oscillating term as a physical mass, the wave function of neutrino mass eigenstates must evolve quasi-adiabatically. This condition is satisfied when $\dot E_\nu/E_\nu^2 = m_i \dot m_i/E_\nu^3 \ll 1$, which holds for relativistic neutrinos, ensuring the validity of the approximation for the scenario considered here. 

For three generations of neutrinos, the neutrino oscillation probability from flavor $\alpha$ to $\beta$ is given by~\cite{ParticleDataGroup:2024cfk}
\begin{align}
\label{osc_probability}
P_{\alpha\beta} &= \delta_{\alpha\beta} - 4\sum_{i<j}^3 \mathrm{Re}\{U_{\alpha_i} U_{\beta i}^* U_{\alpha j}^* U_{\beta j}\} F_{ij} \nonumber\\
&\phantom{=} + 2 \sum_{i<j}^3 \mathrm{Im}\{U_{\alpha_i} U_{\beta i}^* U_{\alpha j}^* U_{\beta j}\} G_{ij} ~,
\end{align}
where $U$ is the Pontecorvo-Maki-Nakagawa-Sakata (PMNS) matrix, $X_{ij} \equiv \Delta m_{ij}^2 L/(4 E_\nu)$, and
\begin{align}
F_{ij} = \sin^2 X_{ij} ~,\quad
G_{ij} = \sin(2 X_{ij}) ~.
\end{align}
The oscillation probability for antineutrinos follows a similar expression, with the substitution $U\rightarrow U^*$. According to \eqref{mass_squared_difference}, the neutrino mass-squared difference varies periodically in time with the period $T_\phi = \pi/m_\phi = 5.7 \,\mathrm{hr}\,(10^{-19}\,\mathrm{eV}/ m_\phi)$. Given the lower bound on the dark matter mass, $m_\phi \gtrsim 3\times 10^{-19}\,\mathrm{eV}$~\cite{Dalal:2022rmp, Amin:2022nlh}, this period is much shorter than the typical timescales of oscillation experiments. For instance, the Jiangmen Underground Neutrino Observatory (JUNO) experiment aims for sub-percent precision in measuring neutrino mass-squared differences within $\sim 100$ days~\cite{JUNO:2022mxj, JUNO:2015zny}. Therefore, we consider the time-averaged probability, which is given by modifying \eqref{osc_probability} with
\begin{align}
\label{fij_timeaverage}
F_{ij} &= \frac{1}{2}[1 - J_0(X_{ijD}) \cos(X_{ijD})] ~,\\
\label{gij_timeaverage}
G_{ij} &= J_0(X_{ijD}) \sin(X_{ijD}) ~,
\end{align}
where $J_n(z)$ is the Bessel function of the first kind and $X_{ijD} \equiv \Delta m_{ijD}^2(\boldsymbol x) L/(4 E_\nu)$. These modifications lead to distorted oscillation probabilities, which we will use to compare with oscillation data.

{\bf \emph{Impacts of dark matter fluctuations.}}---Apart from time modulation, the mass-squared difference \eqref{mass_squared_difference} also depends on local variations in dark matter density. The typical coherence length is set by the de Broglie wavelength of dark matter,
\begin{align}
\label{de_broglie_length}
\lambda_\mathrm{dB} = \frac{2\pi}{m_\phi v} = 1.24\,\mathrm{au} \left( \frac{10^{-14}\,\mathrm{eV}}{m_\phi} \right) \left( \frac{200\,\mathrm{km/s}}{v} \right) ~.
\end{align}
During the course of an experiment, the Earth's displacement is
\begin{align}
\label{crossing_distance}
l_\oplus = T_\mathrm{exp} v_\oplus = 1.16\,\mathrm{au} \left( \frac{T_\mathrm{exp}}{10\,\mathrm{days}} \right) \left( \frac{v_\oplus}{200\,\mathrm{km/s}} \right) ~,
\end{align}
with $T_\mathrm{exp}$ is the duration of the experiment and $v_\oplus$ is Earth's speed. The amplitude of the mass-squared difference $\Delta m_{ijD}^2$ can be treated as constant during the experiment if the coherence length \eqref{de_broglie_length} exceeds the crossing distance of the Earth \eqref{crossing_distance}. Therefore, the time averaged probability, given by \eqref{osc_probability} with \eqref{fij_timeaverage} and \eqref{gij_timeaverage}, is applicable to oscillation experiments if the dark matter mass is $m_\phi\ll 10^{-14}\,\mathrm{eV}$.

For cases where $m_\phi \gg 10^{-14}\,\mathrm{eV}$, the Earth traverses regions with different scalar field amplitudes during the experiment. To compare with data, the oscillation probability must be spatially averaged once the spatial distribution of dark matter is understood. On galactic scales, the dark matter density is well described by the Navarro-Frenk-White profile~\cite{Navarro:1996gj}, and the inferred value at the solar neighborhood is $\rho_\mathrm{loc} \simeq 0.4\,\mathrm{GeV/cm^3}$~\cite{Salucci:2010qr, Pato:2015dua, Bovy:2012tw}. On subgalactic scales, however, there could be large ($\gtrsim \mathcal O(1)$) deviations from $\rho_\mathrm{loc}$ due to constructive and destructive interference of the dark matter field~\cite{Nakatsuka:2022gaf, Centers:2019dyn, Lisanti:2021vij, Eggemeier:2022hqa}. The detailed distribution depends on several factors, including the dark matter mass, interaction types, halo properties, and substructure evolution (see~\cite{Eggemeier:2022hqa, Chen:2024pyr, Dalal:2020mjw, Chen:2020cef} for examples).

\begin{figure}
\centering
\includegraphics[width=0.9\linewidth]{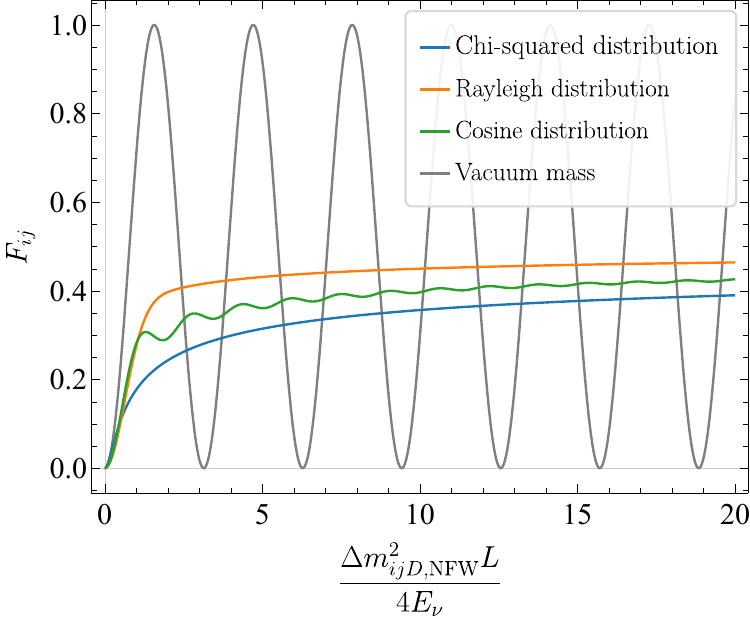}
\caption{Values of $F_{ij}$, defined in the flavor-changing probability \eqref{osc_probability}, as a function of $\Delta m_{ijD, \mathrm{NFW}}^2 L/(4 E_\nu)$, where $\Delta m_{ijD, \mathrm{NFW}}^2$ denotes the amplitude inferred from an NFW profile. Different assumptions for dark matter density fluctuations are considered, see text for details. While quantitative differences arise, dark matter fluctuations generally suppress the oscillatory behavior of spacetime-averaged flavor-changing probabilities as a function of the neutrino propagating distance $L$. For reference, the gray line shows the result corresponding to vacuum neutrino mass, for which $\Delta m_{ijD, \mathrm{NFW}}^2$ is understood to be independent of dark matter density.}
\label{fig:fij}
\end{figure}

For definiteness, we assume that the dark matter field values are homogeneous and uncorrelated across different de Broglie patches and follow a Gaussian distribution at a fixed radius $R$ from the Galactic center,
\begin{align}
\label{field_distribution}
\frac{\phi(\boldsymbol x)}{\phi_\mathrm{NFW}(r)} \Big|_R \sim N(0,1) ~,
\end{align}
where $\phi_\mathrm{NFW}$ is the value corresponding to the NFW profile, i.e., $\rho_\mathrm{NFW}(r) = m_\phi^2 \phi_\mathrm{NFW}^2(r)/2$. According to \eqref{field_distribution}, the dark matter density follows a chi-squared distribution, $\rho(\boldsymbol x)/\rho_\mathrm{NFW}(r) \sim \chi^2(1)$. To compute the spatial average, we begin by performing a Taylor expansion of the time-averaged expressions in \eqref{fij_timeaverage} and \eqref{gij_timeaverage}, yielding polynomial forms in terms of $\Delta m_{ijD}^2(\boldsymbol x)$. Assuming that the spatial average equals the ensemble average, we then evaluate the spatial average of each polynomial term by noting that $\Delta m_{ijD}^2\propto \rho(\boldsymbol x)$ follows the same chi-squared distribution. The final result is obtained by summing over all polynomial contributions. The behavior of the time- and space-averaged quantity $F_{ij}$ is illustrated in Fig.~\ref{fig:fij}. As shown, the neutrino oscillation probability ceases to be an oscillatory function of distance. Given the observed oscillation behavior, we conclude that dark matter with $m_\phi \gg 10^{-14}\,\mathrm{eV}$ cannot account for neutrino masses.

The robustness of the conclusion derives from its insensitiveness to the detailed distribution of dark matter density fluctuations. In addition to the chi-squared distribution, Fig.~\ref{fig:fij} also shows $F_{ij}$ for the case where $\rho(\boldsymbol x)/\rho_\mathrm{NFW}(r)$ follows the Rayleigh distribution or varies periodically as $1-\cos(k_\theta \theta)$, where $\theta$ is the azimuthal angle of the Earth with respect to the Galactic center and $k_\theta$ is a constant satisfying $k_\theta\gg R/\lambda_\mathrm{dB}$. For comparison, we also include the result for constant $\rho(\boldsymbol x)$, which leads to constant neutrino masses during oscillation experiments. Despite the quantitative difference, in all three distributions the oscillatory behaviors in terms of $X_{ijDR}$, and thus the neutrino propagating distance $L$, is significantly suppressed.

{\bf \emph{Tests with neutrino oscillation data.---}}
We now explore the compatibility between the distorted oscillation probability, i.e., Eqs.~\eqref{osc_probability}, \eqref{fij_timeaverage}, and~\eqref{gij_timeaverage}, and neutrino oscillation data. We refer to the ``dark mass hypothesis'' as the model investigated here, in which the active neutrino acquires its small mass through interactions with the dark matter field. For short baseline reactor experiments such as RENO~\cite{RENO:2018dro}, Double Chooz~\cite{DoubleChooz:2019qbj} and Daya Bay~\cite{DayaBay:2022orm}, the electron survival probability is mainly determined by $|\Delta m_{32}^2|/|\Delta m_{31}^2|$ and the $\theta_{13}$ mixing angle. By performing a $\chi^2$ analysis using data from~\cite{RENO:2018dro,DayaBay:2022orm}, we reconstruct the parameters $\Delta m_{32}^2$, $\Delta m_{32D}^2$ and $\theta_{13}$, and find that the neutrino dark mass hypothesis provides a good fit. In order to make comparison between the vacuum and dark mass hypotheses we convert the differences in their respective $\chi^2_{\rm min}$ into a significance level~$\sigma$. We do this with a two-sided p-value because we want to remain agnostic about which model provides a better fit to the data. Typically, when one compares a new model with the null hypothesis, one uses a one-sided p-value. Instead, here we treat the two models as similarly complex alternatives, using a two-sided test to detect any significant difference in fit. Since our datasets and number of free parameters are the same for each model, the degrees of freedom used in converting $\Delta\chi^2$ to $\sigma$ is simply the number of parameters fitted for. The best-fit results from Daya Bay favor the vacuum hypothesis by a statistical significance of only $1.7\sigma$, while the RENO data yield a preference for the dark mass hypothesis, although the significance is relatively modest at $\sim0.5\sigma$.

For long baseline experiments, we consider the results from the Kamioka Liquid Scintillator Antineutrino Detector (KamLAND) experiment~\cite{KamLAND:2013rgu}, where the survival probability is primarily influenced by $\Delta m_{21}^2$, $\theta_{12}$ and $\theta_{13}$. In our analysis, we use the binned survival probability data as presented in Fig.~5 of~\cite{KamLAND:2013rgu}. This approach has the advantage of avoiding the need to model the backgrounds, because we account for the background-subtracted result of the collaboration. We do not consider background and systematic uncertainties, which are expected to be small for high energy bins $\gtrsim 3\,\mathrm{MeV}$~\cite{KamLAND:2013rgu}. To simulate the data, we first compute a flux-weighted survival probability $\sum_i \Phi_i P_{ee}(L_i/E_\nu)/\sum_i \Phi_i$, where the sum runs over all reactors at distance $L_i$, $\Phi_i$ is the total neutrino flux received at KamLAND for each reactor, and the flux values are collected from Fig.~32 of~\cite{Bemporad:2001qy}. We then average over the 20 bins to match the data
\begin{equation}
    \langle \bar{P}_{ee}
\rangle = \frac{1}{\Delta x}\int\,{\rm d}x\, \frac{\sum_i\Phi_i P_{ee}(L_i/E_\nu)}{\sum_i \Phi_i}
\end{equation}
where, once the flux-weighted average reactor baseline is fixed as $L=180\,\mathrm{km}$~\cite{KamLAND:2013rgu}, we define $x\equiv L/E_\nu$ and the bin width $\Delta x$. The integral is approximated by an unweighted average within each bin. For parameter fitting, we employ the nested sampling algorithms implemented in the UltraNest code~\cite{2016S&C....26..383B, 2019PASP..131j8005B, 2021JOSS....6.3001B}. We first reconstruct the parameters $\left\{\Delta m_{21}^2, \theta_{12}, \theta_{13}\right\}$ using only KamLAND data, considering both the vacuum and dark mass hypotheses. In both cases, we adopt the flat prior $\Delta m_{21}^2 \in \left[1-15\right]\times 10^{-5}\,{\rm eV}^2$, while the mixing angles $\theta_{12}$ and $\theta_{13}$ vary within $\left[0-1\right]$. 

\begin{figure}
    \centering
    \includegraphics[width=1.0\linewidth]{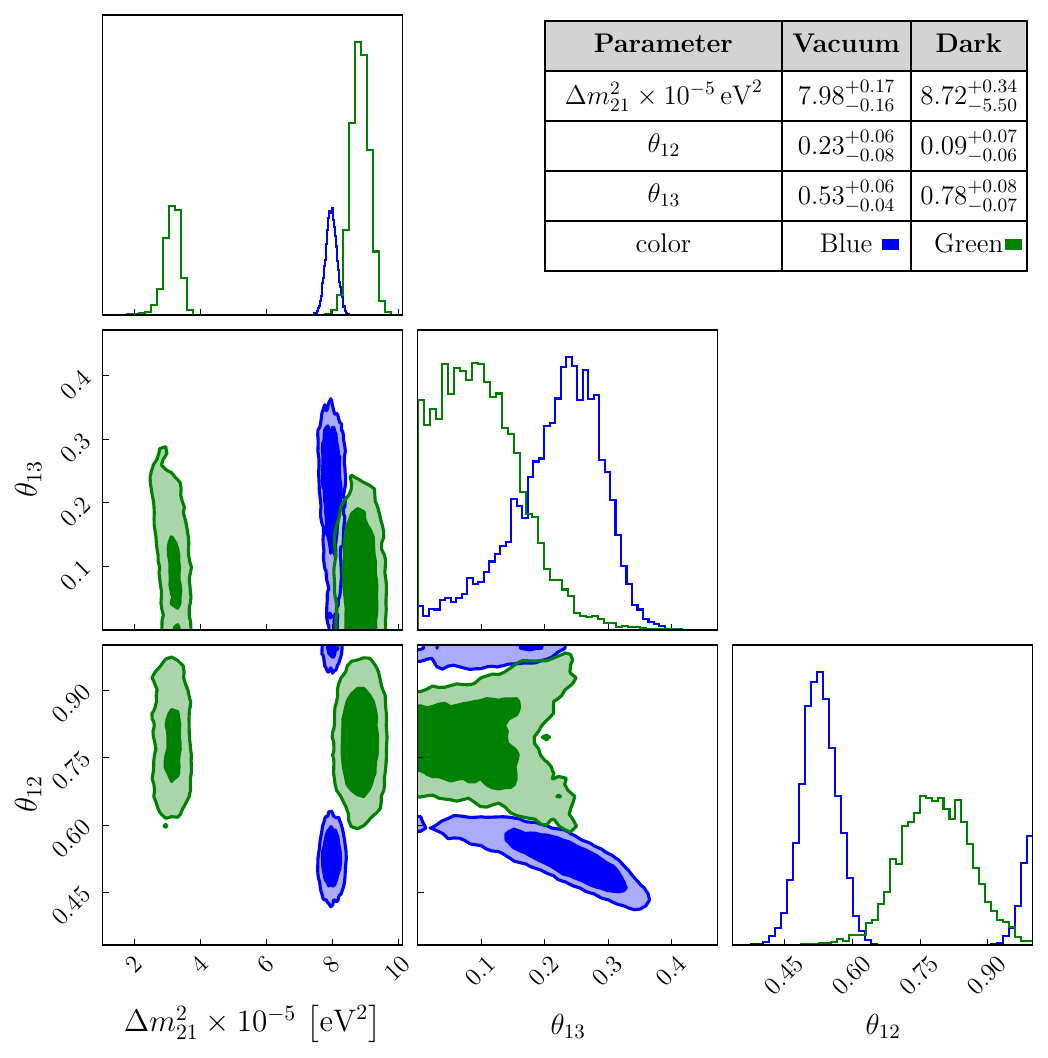}
    \caption{Parameter fits discussed in the main text. Results are shown for the 3-parameter fit $\left\{\Delta m_{21}^2, \theta_{12}, \theta_{13}\right\}$ using only KamLAND data. Contours represent the 68\% and 95\% confidence levels for the vacuum (blue) and dark (green) neutrino mass hypotheses, respectively.}
    \label{fig:param_scan3D}
\end{figure}

The parameter reconstructions for each scan are shown in Fig.~\ref{fig:param_scan3D}. We present the results for both the vacuum mass (blue shaded) and dark mass (green shaded) hypothesis in the same plot, with contours corresponding to 68\% and 95\% confidence levels. By showing both models together, we highlight the regions of parameter space favored by each scan. However, since the probabilities are normalized to the best-fit point of each respective scan, this plot does not indicate which hypothesis provides a better fit. For this, we compare the minimum $\chi^2$ value of each hypothesis and find that the vacuum mass hypothesis is favored at the $4.5\sigma$ level, with $\chi^2_{\rm min,\, dark}=61.7$ and $\chi^2_{\rm min,\, vac}=35.5$. Our best-fit $\chi^2$ is slightly larger than the value inferred from Fig.~5 of Ref.~\cite{KamLAND:2013rgu}, $\chi^2\approx 33$, likely due to the simplified nature of our analysis and the lack of direct access to raw data. However, we do not expect this difference to significantly alleviate the tension between the models. Consequently, our finding that KamLAND alone imposes serious constraints of the neutrino dark mass hypothesis remains robust.

Notice that for the dark mass hypothesis there are two separate regions of parameter space which constitute its best-fit, at $\Delta m_{21D}^2\approx 3\times 10^{-5}\,{\rm eV}^2$ (Dark 1) and $\Delta m_{21D}^2\approx 9\times 10^{-5}\,{\rm eV}^2$ (Dark 2). This arises from the oscillatory behavior of the time-averaged probability given in Eq. \eqref{fij_timeaverage}. Additionally, an alternative best-fit region appears for the vacuum mass hypothesis at $\theta_{12} > 0.9$, which is of lesser interest given that solar neutrino experiments determine $\theta^\mathrm{solar}_{12} \approx 0.59$ \cite{Esteban:2024eli}.

\begin{figure}
    \centering
        \includegraphics[width=1.00\linewidth]{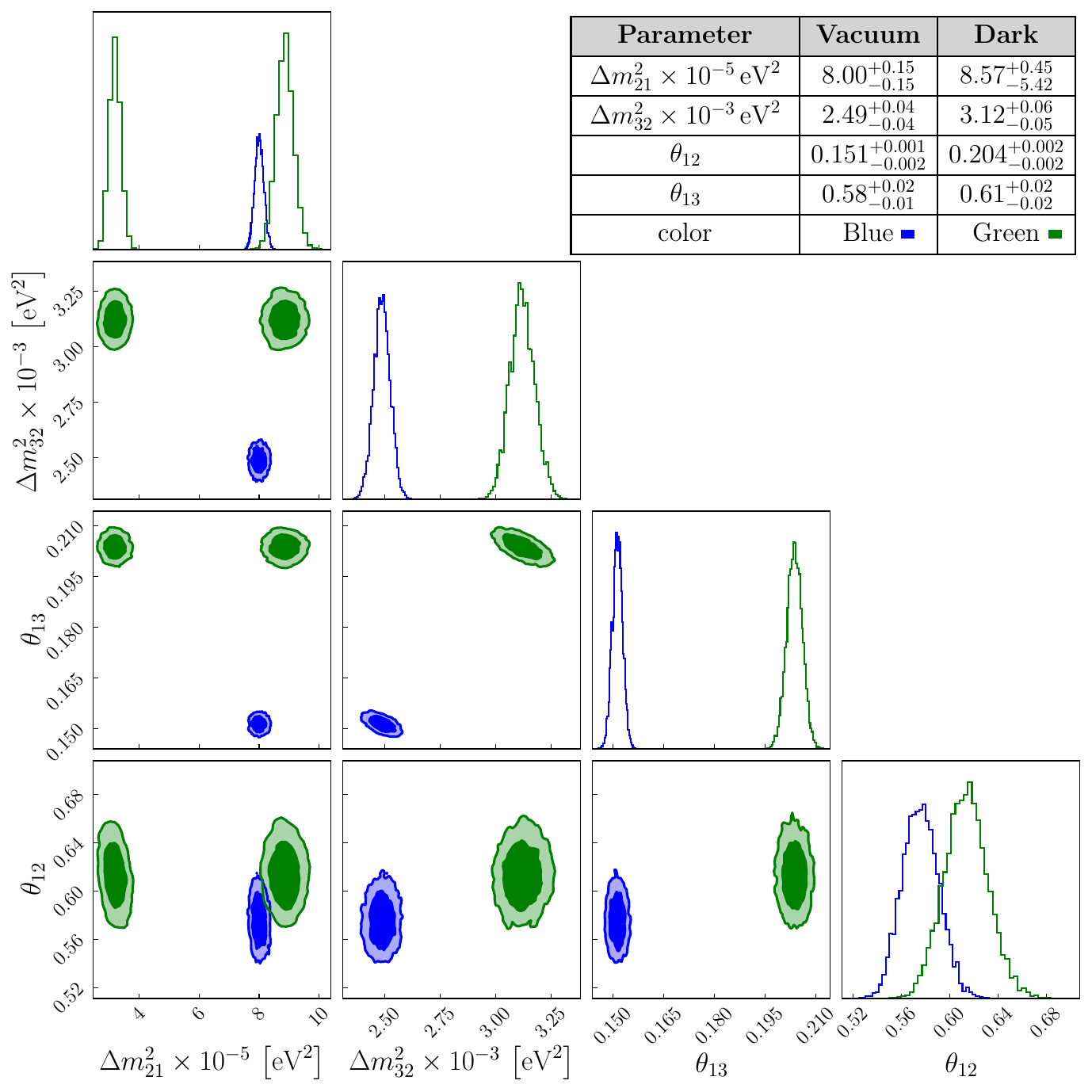}
    \caption{Same as Fig.~\ref{fig:param_scan3D}, but for the 4-parameter fit $\left\{\Delta m_{32}^2, \Delta m_{21}^2, \theta_{12}, \theta_{13}\right\}$ obtained from a combined analysis of KamLAND, RENO, Double Chooz data, including $\theta_{12}^{\rm solar}$ as a pull parameter. }
    \label{fig:param_scan4D}
\end{figure}

For Fig.~\ref{fig:param_scan4D} we combine short baseline and KamLAND data with the solar determination of $\theta_{12}$ to perform a 4D parameter reconstruction in $\left\{\Delta m_{21}^2,\Delta m_{32}^2, \theta_{12}, \theta_{13}\right\}$. The cleanest result to incorporate into our analysis is for solar neutrinos. This is because $\theta_{12}$ is the dominant parameter determining the physics~\cite{ParticleDataGroup:2024cfk}. We combine the solar determination of $\theta_{12}$ by introducing it as a nuisance parameter in the $\chi^2$ determination,
\begin{equation}
    \chi^2 =  \frac{\left(\theta_{12}-\theta^{\rm solar}_{12}\right)^2}{\sigma^2_{\rm solar}}+ \sum_{\rm exps}\chi^2 , \label{eq:ball_param}
\end{equation}
where the central value $\theta_{12}^{\rm solar}\approx0.59$ and uncertainty $\sigma_{\rm solar}\approx 0.025$ are taken from Fig.~11 of~\cite{Esteban:2024eli}.\footnote{Note that a more rigorous treatment of the $\theta_{12}$ effect would involve recomputing the solar neutrino fit for each tested value of $\delta m^2_{12}$, since small correlations do exist, e.g.\ through the mixing in the resonance region. However, the published global solar+KamLAND results show that $\sin^2\theta_{12}$ remains stable at the level of $0.01$ even when $\delta m^2_{12}$ shifts by its current uncertainty of $\sim 1\sigma$~\cite{Bahcall:2004ut}. This subtle dependence has a small numerical impact on our analysis. A fully consistent solar reanalysis is beyond the scope of this study.} Our KamLAND-only analysis, shown in Fig.~\ref{fig:param_scan3D}, indicates that the neutrino mass hypothesis favors $\theta_{12}^{\rm dark}= 0.78\pm^{0.08}_{0.07}$, which is in tension with $\theta^{\rm solar}_{12}$. In contrast, for the vacuum mass hypothesis, $\theta_{12}^{\rm vac}=0.53\pm^{0.06}_{0.04}$, which is consistent with $\theta^{\rm solar}_{12}$.

\begin{figure}
\centering
\includegraphics[width=\linewidth]{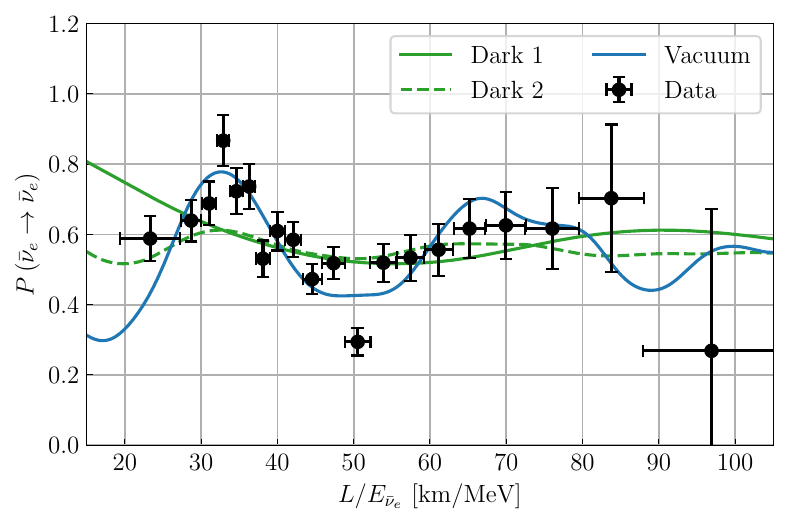}
\caption{The $\bar{\nu}_e$ survival probability as a function of $L/E_{\bar{\nu}_e}$ for the KamLAND experiment. The black points represent the binned data as reported in~\cite{KamLAND:2013rgu} and the lines correspond to the unbinned survival probabilities with the best-fit parameters for the vacuum or dark mass hypothesis respectively. For the dark mass hypothesis, there are two sets of best-fit parameters corresponding to two peaks in Fig. \ref{fig:param_scan4D}, denoted as Dark 1 and Dark 2. Here $L=180\,\mathrm{km}$ is the flux-weighted average reactor baseline.}
\label{fig:BP4D_signal}
\end{figure}

The confidence intervals for this parameter reconstruction are shown in Fig.~\ref{fig:param_scan4D}. While we observe a narrowing of the best fit parameters, the overall picture is similar to the results of the 3D scan. Once again, we compare the two models by computing the minimum $\chi^2$ values. The neutrino dark mass hypothesis is now disfavored at the $4.9\sigma$ level, with $\chi^2_{\rm min,\, dark}=89.9$ and $\chi^2_{\rm min,\, vac}=56.5$. The reduced ability of the dark mass hypothesis to fit the data is driven by the solar value of $\theta_{12}$, while short baseline experiments can still be accommodated relatively well. In Fig.~\ref{fig:BP4D_signal} we present the best-fit survival probability for the 4D scan for the vacuum case (``Vacuum'') and the dark mass hypothesis, which includes the two best-fit regions, Dark 1 and Dark 2, obtained from the analysis in Fig.~\ref{fig:param_scan4D}. The jagged structure of the vacuum mass oscillation (blue) curve is a consequence of the flux-weighted average over the reactors from which antineutrinos are detected by KamLAND~\cite{KamLAND:2013rgu}.

A global analysis with all neutrino experiments would provide a more robust comparison between the vacuum and neutrino dark mass origin hypotheses, and we leave this for future work. What our analysis above already shows however is that the vacuum neutrino mass is likely favored to a statistically significant level. Additionally as discussed above, the JUNO experiment expects to determine $\Delta m_{21}^2$ and $\Delta m_{31}^2$ by a relative precision at the sub-percent level within $\sim 100$\,days. By extrapolating this expectation to lower precision, we estimate that a $10\%$ precision can be achieved after $\sim 10\,{\rm days}$. Thus, if occurring in a time scale from 10 days to a few decades, the change of oscillation parameters due to dark matter density fluctuations will be detectable in current and future experiments. According to \eqref{de_broglie_length} and \eqref{crossing_distance}, this corresponds to $10^{-17}\,\mathrm{eV}\lesssim m_\phi\lesssim 10^{-14}\,\mathrm{eV}$.

{\bf \emph{Discussions and Conclusion.---}}
In this work, we consider neutrino interactions with scalar dark matter as the origin of neutrino masses~\cite{Ge:2018uhz, Choi:2019zxy, Sen:2023uga, Capozzi:2018bps, Lee:2024rdc, Huang:2022wmz, Plestid:2024kyy, ChoeJo:2023ffp}. This scenario gives rise to intriguing phenomenology, including oscillations between the Dirac and Majorana nature of neutrinos~\cite{ChoeJo:2023ffp} and a distinctive time delay pattern of supernova neutrinos~\cite{Ge:2024ftz}. To assess its compatibility with neutrino oscillation data, we derive the oscillation probabilities in this scenario by incorporating the effects of temporal oscillations and spatial variations in the dark matter field. 

For $m_\phi\gg 10^{-14}\,\mathrm{eV}$, the Earth's motion through regions with varying scalar field amplitudes induces fluctuations in the neutrino masses over the duration of an experiment. The spatial (along with the temporal) average of the oscillation probability significantly suppresses its oscillating behavior over neutrino propagating distances, regardless of the detailed distribution of dark matter profiles. For $m_\phi\ll 10^{-14}\,\mathrm{eV}$, rapid temporal variations in the field configuration distort the oscillation probabilities, altering the predicted neutrino masses. Comparing with oscillation data, our $\chi^2$ analysis reveals a strong tension, exceeding $4\sigma$, between the time-averaged oscillation probability and data from reactor neutrino experiments, including RENO, Double Chooz, Daya Bay, and KamLAND. More specifically, when assessing the parameters $\left\{\Delta m_{21}^2, \theta_{12}, \theta_{13}\right\}$ against KamLAND data alone, the vacuum mass hypothesis is favored over the dark mass hypothesis at the $4.5\sigma$ level. This preference increases to $4.9\sigma$ when the analysis is extended to include $\Delta m_{32}^2$ and short baseline reactor data. We thus conclude that neutrino interactions with scalar dark matter, as proposed in~\cite{Choi:2019zxy, Sen:2023uga, Capozzi:2018bps, Lee:2024rdc, Huang:2022wmz, Plestid:2024kyy, ChoeJo:2023ffp}, are unlikely to be the dominant mechanism behind neutrino mass generation.

Due to the limited access to the KamLAND data, our analysis relies on more assumptions than the collaboration paper~\cite{KamLAND:2013rgu}, including the omission of systematic errors, the use of an unweighted average in binned probabilities, and the reliance on reactor information published in 2002~\cite{Bemporad:2001qy}. To mitigate the impact, we have verified that our conclusions are qualitatively unchanged even when accounting for the bin width of data as uncertainties in $x$-axis, which increases the tolerance to different models. Furthermore, the dark matter mass range $10^{-17}\,\mathrm{eV} \lesssim m_\phi \lesssim \,10^{-14}\,\mathrm{eV}$ can be further tested in current and future oscillation experiments by searching for time variations in oscillation parameters, which is due to the stochastic nature of dark matter density fluctuations and are, in general, not periodic \cite{Nakatsuka:2022gaf, Centers:2019dyn, Lisanti:2021vij}.

\begin{acknowledgments}
We would especially like to thank Chui-Fan Kong for insightful discussions. We would also like to thank Dorian Amaral, Stefano Gariazzo, Gaetano Lambiase, Andrew Long, Iwan Morton-Blake, Yuber Perez-Gonzalez, Shun Zhou, and Tailin Zhu for helpful comments. AC is supported by the National Natural Science Foundation of China (NSFC) through the grants No.12425506, 12375101, 12090060, and 12090064, and the SJTU Double First Class start-up fund (WF220442604). LV and HYZ acknowledge support by NSFC through the grant No. 12350610240 ``Astrophysical Axion Laboratories''. LV also thanks INFN through the ``QGSKY'' Iniziativa Specifica project, as well as the COST Actions ``COSMIC WISPers'' (CA21106) and ``Addressing observational tensions in cosmology with systematics and fundamental physics (CosmoVerse)'' (CA21136), both supported by COST (European Cooperation in Science and Technology).
\end{acknowledgments}

\bibliography{ref}

\end{document}